%% file: main.tex
\documentclass[%
 reprint,
 superscriptaddress,
 amsmath,amssymb,
 aps,
]{revtex4-2}

\usepackage{graphicx}
\usepackage{bm}
\usepackage{siunitx}
\usepackage{hyperref}
\usepackage{physics}
\usepackage{placeins}

\DeclareSIUnit[]{\dBm}{dBm}

\newcommand{\cofirstauthorinfo}{%
  \thanks{These authors contributed equally to this work}%
}

\newcommand{\correspondingathorinfo}{%
  \thanks{Corresponding author}%
  \email{kasper.vangasse@ugent.be}%
}

\begin{document}

\title{Integrated lasing and dual-band gain in neodymium-implanted silicon nitride}


\author{Jan de Haan}
\cofirstauthorinfo
\affiliation{Photonics Research Group, INTEC, Ghent University - imec, 9052 Ghent, Belgium}
\affiliation{Center for Nano and Biophotonics (NB-Photonics), Ghent University, Technologiepark-Zwijnaarde 126, 9052 Ghent, Belgium}

\author{Arturo Fernández Gamez}
\cofirstauthorinfo
\affiliation{Photonics Research Group, INTEC, Ghent University - imec, 9052 Ghent, Belgium}
\affiliation{Center for Nano and Biophotonics (NB-Photonics), Ghent University, Technologiepark-Zwijnaarde 126, 9052 Ghent, Belgium}

\author{Nina Zdravković}
\affiliation{Photonics Research Group, INTEC, Ghent University - imec, 9052 Ghent, Belgium}
\affiliation{Center for Nano and Biophotonics (NB-Photonics), Ghent University, Technologiepark-Zwijnaarde 126, 9052 Ghent, Belgium}

\author{Kevin Kwock}
\affiliation{Photonics Research Group, INTEC, Ghent University - imec, 9052 Ghent, Belgium}
\affiliation{Center for Nano and Biophotonics (NB-Photonics), Ghent University, Technologiepark-Zwijnaarde 126, 9052 Ghent, Belgium}

\author{Ulrich Kentsch}
\affiliation{Helmholtz-Zentrum Dresden-Rossendorf (HZDR), Dresden, Germany}

\author{Rene Heller}
\affiliation{Helmholtz-Zentrum Dresden-Rossendorf (HZDR), Dresden, Germany}

\author{Alberto Curto}
\affiliation{Photonics Research Group, INTEC, Ghent University - imec, 9052 Ghent, Belgium}
\affiliation{Center for Nano and Biophotonics (NB-Photonics), Ghent University, Technologiepark-Zwijnaarde 126, 9052 Ghent, Belgium}

\author{Bart Kuyken}
\affiliation{Photonics Research Group, INTEC, Ghent University - imec, 9052 Ghent, Belgium}
\affiliation{Center for Nano and Biophotonics (NB-Photonics), Ghent University, Technologiepark-Zwijnaarde 126, 9052 Ghent, Belgium}

\author{Kasper Van Gasse}
\correspondingathorinfo
\affiliation{Photonics Research Group, INTEC, Ghent University - imec, 9052 Ghent, Belgium}
\affiliation{Center for Nano and Biophotonics (NB-Photonics), Ghent University, Technologiepark-Zwijnaarde 126, 9052 Ghent, Belgium}

\date{\today}

\begin{abstract}
Lasers at wavelengths near \SI{1.0}{\um} are used in quantum technologies, precision metrology, and sensing. Rare-earth gain at these wavelengths in photonic integrated circuits can reduce the size and enable the scalable fabrication of these lasers. We fabricate neodymium-based integrated waveguide amplifiers and a ring laser using neodymium-implanted LPCVD silicon nitride. The photonic integrated circuits were fabricated on a 200 mm commercial wafer platform, and individual dies were implanted with neodymium ions. We show that the amplifiers provide gain in two bands around wavelengths of \SI{920}{\nm} and \SI{1100}{\nm}, and we show lasing in a ring resonator. This demonstrates that neodymium-implanted LPCVD silicon nitride is a suitable gain material for integrated lasers.
\end{abstract}

\maketitle
\section{Introduction}

Since the first demonstrations of neodymium-doped glass (Nd:glass) lasers in 1961 and neodymium-doped yttrium aluminum garnet (Nd:YAG) lasers in 1964 \cite{Snitzer1961NdGlassLaser,VanUitert1964NdYAGLaser},
Nd$^{3+}$-based gain media have played a central role in solid-state laser technology. Both systems exploit the $^4F_{3/2} \rightarrow {}^4I_{11/2}$ transition of Nd$^{3+}$, providing near-infrared emission at a wavelength around \SI{1.06}{\um}, typically under optical pumping at a wavelength near \SI{808}{\nm}. Nd:YAG combines a large emission cross section, favorable four-level laser dynamics, and excellent thermal properties, making it particularly suitable for efficient continuous-wave and high-power operation. Nd:glass, on the other hand, offers a broader emission bandwidth, favorable energy-storage properties, and greater flexibility in material composition and fabrication, establishing its importance in high-energy laser
amplification and ultrashort-pulse systems
\cite{Snitzer1966GlassLaserReview,Suratwala2000NdGlassReview}.

\begin{figure*}[t]
    \centering
    \includegraphics{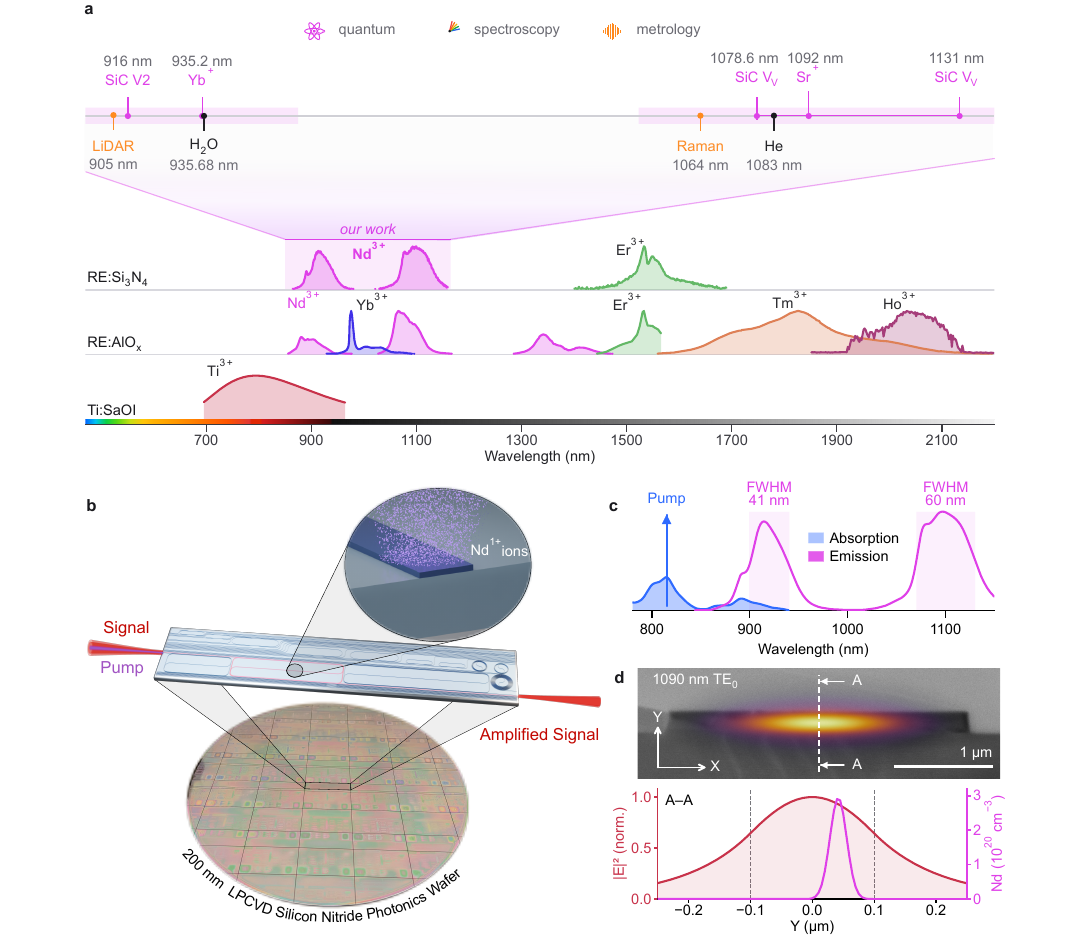}
\caption{%
  Overview of Nd$^{3+}$:Si$_3$N$_4$ on-chip amplifiers
  \textbf{(a)}~Normalized emission spectra grouped by host. Data from: Er:SiN~\cite{Kippenberg2022ErAmplifiers} (deposited data)., Nd:AlO$_x$~\cite{Pollnau2010AlOxNd}, Yb:AlO$_x$~\cite{Pollnau2025YbAlumina}, Er:AlO$_x$~\cite{Smit1997ErAlOx}, Tm:AlO$_x$~\cite{Kaertner2025TmAlumina}, Ho:AlO$_x$~\cite{Watts2018HoAlumina} (digitized data), Ti:SaOI (fitted gain)~\cite{Vuckovic2024TiSaOI}. 
  \textbf{(b)}~Commercial 200~mm silicon nitride wafer and Nd-implanted Si$_3$N$_4$  spiral amplifier die.
  \textbf{(c)} Schematic of the Nd$^{3+}$ absorption bands and emission bands in Si$_3$N$_4$ under optical pumping at \SI{815}{\nm}.
  \textbf{(d)}~Top panel: Scanning electron microscope image of an unimplanted waveguide overlaid with the simulated TE$_0$ mode at a wavelength of \SI{1090}{\nm}. Bottom panel: Normalized field intensity and simulated Nd concentration along the line marked A--A in the top panel.
}
    \label{fig:1-overview-and-fabrication}
\end{figure*}

Despite their widespread use in industrial, scientific, and medical applications, conventional Nd-based laser systems often rely on bulk gain media and free-space optical components. Nd-doped fiber lasers offer a more compact alternative, benefiting from efficient optical confinement and improved thermal management. Meanwhile, the growing demand for compact and scalable photonic systems, largely driven by the telecom and datacom markets, has accelerated the development of silicon photonics, combining low-loss passive waveguides, high-speed modulators, and heterogeneously integrated light sources. In this context, rare-earth-doped waveguide amplifiers and lasers have emerged as promising candidates for integrating optical gain on-chip. In particular, erbium-doped waveguide amplifiers (EDWAs), the integrated counterparts of erbium-doped fiber amplifiers (EDFAs), have been demonstrated through ion implantation into low-loss silicon nitride ($\mathrm{Si}_3\mathrm{N}_4$) waveguides \cite{Kippenberg2022ErAmplifiers,Kippenberg2026ThinNitrideLasers}. In parallel, reactive co-sputtering of rare-earth-doped aluminum oxide ($\mathrm{Al}_2\mathrm{O}_3$) has enabled not only erbium-based integrated amplifiers and lasers but also other rare-earths such as thulium (Tm$^{3+}$), and holmium (Ho$^{3+}$), operating across different wavelength bands \cite{GarciaBlanco2025ErAlOxPackaged,Kaertner2025TmAlumina,Watts2018HoAlumina}. These developments (Fig.~\ref{fig:1-overview-and-fabrication} (a)) highlight the versatility of rare-earth-doped materials for integrated optical gain, while the combination of low-loss silicon nitride photonics and ion implantation, both well-established in CMOS fabrication, offers a scalable, wafer-level approach to their integration.

In contrast, the integration of Nd$^{3+}$ into thin-film photonic platforms has predominantly relied on sputter-deposition techniques. Reactively co-sputtered Nd-doped aluminum oxide ($\mathrm{Al}_2\mathrm{O}_3:\mathrm{Nd}^{3+}$) has enabled on-chip optical amplification \cite{Pollnau2010AlOxNd,GarciaBlanco2026NdAlOx}, while co-sputtered Nd-doped silicon nitride ($\mathrm{SiN}_x:\mathrm{Nd}^{3+}$) microdisks have demonstrated photoluminescence and stimulated emission, but not lasing \cite{DalNegro2011SputteredNdSiN}. Nd-based microresonator lasing has previously been demonstrated in silica microspheres \cite{Haroche1996MicrosphereLaser}, alumina-containing silica microtoroids \cite{Armani2013NdToroid}, and Nd-doped alumina Taiji resonators \cite{GarciaBlanco2023TaijiResonator}. More recently, integrated lasing has also been achieved by bonding a mm-sized Nd:YAG crystal to a silicon nitride photonic integrated circuit \cite{Tang2026NdYAGMOPA}.

While erbium-based devices primarily target the telecommunication bands around a wavelength of \SI{1.55}{\um}, integrated light sources operating near \SI{1.0}{\um} are of particular interest for quantum technologies, precision spectroscopy, and optical metrology. Examples include solid-state quantum emitters such as divacancy centers in silicon carbide, with optical transitions in the \SI{1.1}{\um} region, as well as frequency-stabilized laser systems for
precision measurements \cite{Awschalom2017SiCQubits,Byer1993NdYAGStabilization}. These applications motivate the development of compact, integrated gain media operating beyond the conventional telecommunication bands.

In this work, we explore, to the best of our knowledge for the first time, Nd$^{3+}$ ion implantation into pre-existing, LPCVD low-loss $\mathrm{Si}_3\mathrm{N}_4$ waveguides  (see Fig.~\ref{fig:1-overview-and-fabrication} (b)). By incorporating optical gain directly into an established photonic platform this approach offers a promising route towards scalable, on-chip optical amplifiers and lasers operating in two gain bands \SI{920}{\nm} and \SI{1100}{\nm} (see Fig.~\ref{fig:1-overview-and-fabrication} (c)) relevant for future quantum systems.

\section{Results}

Low pressure chemical vapor deposition (LPCVD) silicon nitride waveguides with a thickness of \SI{200}{\nm} were implanted with neodymium ions at an energy of \SI{200}{\keV}, resulting in the simulated ion distribution shown in Fig.~\ref{fig:1-overview-and-fabrication}(d).
The waveguides were covered with a \SI{500}{\nm} thick inductively coupled plasma chemical vapor deposition (ICPCVD) silicon oxide film and annealed at \SI{1000}{\celsius} in a nitrogen atmosphere for 1 hour (\hyperref[sec:methods]{Methods}).

We first present measurements of the absorption and emission spectra, as well as a measurement of the lifetime of the ${}^4 F_{3/2}$ level.
Based on the measured lifetime, an upper bound on the emission cross section is given.
Based on the observed absorption and emission spectra, gain measurements are presented in the following section.
From the observed gain, the emission cross section is derived and an estimate of the radiative lifetime is given.
Finally, lasing is demonstrated in a microring resonator.

\subsection{Spectroscopic properties}

\begin{figure*}
\includegraphics{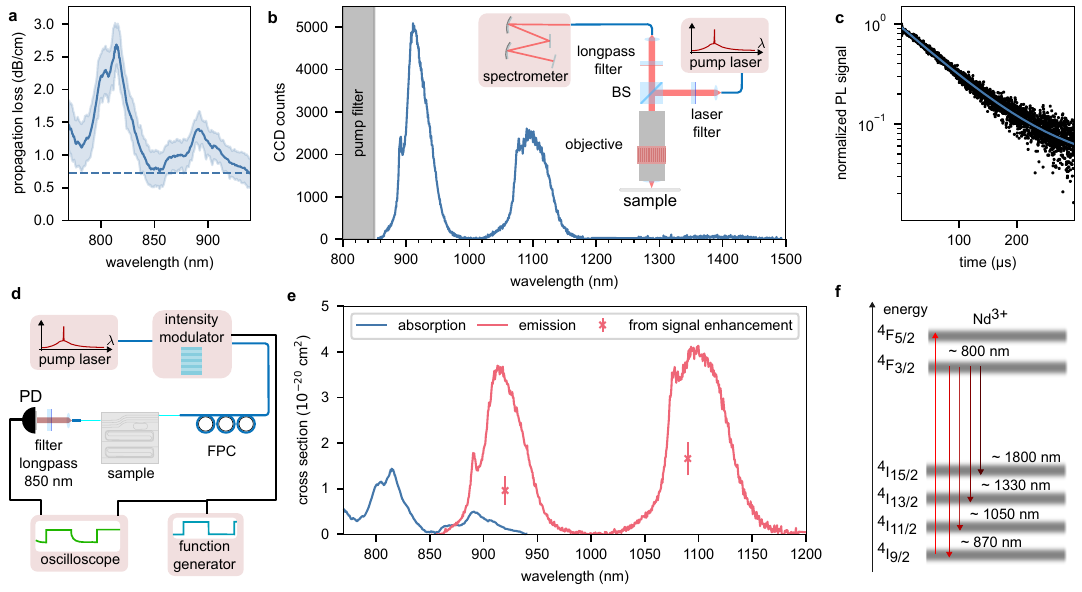}
\caption{\label{fig:2-spectroscopy}\textbf{(a)} Propagation loss spectrum of the Nd-implanted waveguide \textbf{(b)} Emission spectrum of Nd-implanted silicon nitride. Inset: Confocal microscope used to measure the spectrum. BS: Beam splitter \textbf{(c)} Photoluminescence decay with exponential fit with background \textbf{(d)} Setup for photoluminescence decay measurement. FPC: Fiber polarization controller \textbf{(e)} Absorption and emission cross section of $\mathrm{Nd}^{3+}$ in silicon nitride (see text) \textbf{(f)} Energy level diagram for $\mathrm{Nd}^{3+}$ with transition wavelengths \cite{Crosswhite1963RareEarthSpectra}}
\end{figure*}

The propagation loss spectrum measured by the cutback method (see \hyperref[sec:methods]{Methods}) of the waveguides has two maxima at wavelengths around \SI{815}{\nm} and \SI{900}{\nm}, corresponding to the ${}^{4}\mathrm{I}_{9/2} \to {}^{4}\mathrm{F}_{5/2}$ and ${}^{4}\mathrm{I}_{9/2} \to {}^{4}\mathrm{F}_{3/2}$ transitions of $\mathrm{Nd}^{3+}$ (Fig.~\ref{fig:2-spectroscopy}(a) and (f)).

The emission spectrum obtained upon excitation at a wavelength of \SI{815}{\nm} contains two prominent peaks around wavelengths of \SI{920}{\nm} and \SI{1100}{\nm}, corresponding to the ${}^{4}\mathrm{F}_{3/2} \to {}^{4}\mathrm{I}_{9/2}$ and ${}^{4}\mathrm{F}_{3/2} \to {}^{4}\mathrm{I}_{11/2}$ transitions.
There is also some emission slightly above the noise floor observed at \SI{1400}{\nm}, which is assigned to the ${}^{4}\mathrm{F}_{3/2} \to {}^{4}\mathrm{I}_{13/2}$ transition.
The two observed peaks are broad, with full widths at half maximum (FWHM) of \SI{40}{\nm} (\SI{10}{\THz}) for the \SI{920}{\nm} wavelength peak, \SI{60}{\nm} (\SI{15}{\THz}) for the \SI{1100}{\nm} peak.
This large broadening is similar to that in other amorphous hosts (see \hyperref[sec:discussion]{Discussion}).

The lifetime of the ${}^{4}\mathrm{F}_{3/2}$ level is \SI{85}{\us}, shorter than in other amorphous hosts.
The lifetime is measured by the decay of the photoluminescence emitted from the waveguide (Fig.~\ref{fig:2-spectroscopy}(c), (d), and \hyperref[sec:methods]{Methods}).
In aluminum oxide, the lifetime of the ${}^{4}\mathrm{F}_{3/2}$ level is \SI{325}{\us} \cite{Pollnau2010AlOxNd}. In phosphate glasses, its value lies between \SI{300}{\us} and \SI{400}{\us} \cite{Ropp1981NdPhosphateGlass}.

We calculate the emission cross section assuming purely radiative decay of the ${}^{4}\mathrm{F}_{3/2}$ state, i.e. $A = 1 / \tau$. Under this assumption, it can be derived from the measured spectral power in wavelength units $\rho_f(\lambda_0)$ (shown in Fig.~\ref{fig:2-spectroscopy}(b)) of the spontaneously emitted light using the Füchtbauer-Ladenburg relation \cite{Pollnau2015LaserSpectroscopyTheory}
\begin{equation}
    \label{eq:fuchtbauer-ladenburg-relation}
    \sigma_e(\lambda_0) = A \frac{\lambda_0^4}{8 \pi c_0 n^2} \rho_f(\lambda_0).
\end{equation}
The result is shown in Fig.~\ref{fig:2-spectroscopy}(e).

The absorption cross section is calculated from the propagation loss as $\sigma_a(\lambda_0) = \alpha_\mathrm{Nd}(\lambda_0) / (\Gamma(\lambda_p) N_\mathrm{max})$, using the implanted dose and simulated concentration distribution to calculate the number of ions interacting with the mode. It is also shown in Fig.~\ref{fig:2-spectroscopy}(e).

\subsection{Optical amplification at wavelengths of \SI{920}{\nm} and \SI{1090}{\nm}}

\begin{figure*}
\includegraphics{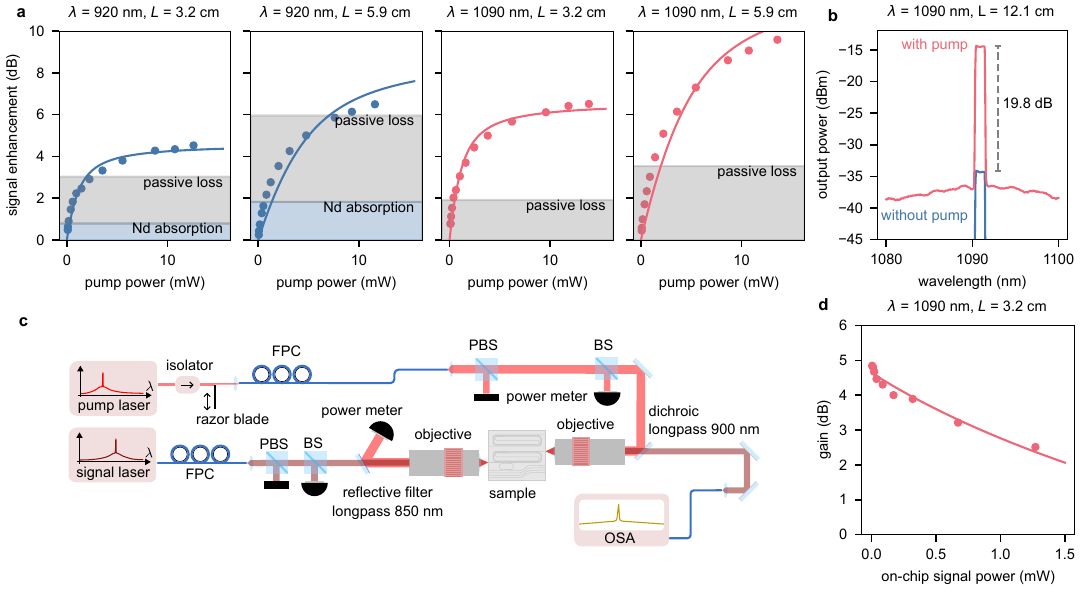}
\caption{\label{fig:3-amplification}\textbf{(a)} Measured signal enhancement at different pump powers for amplifiers of length \SI{3.2}{\cm} and \SI{5.9}{\cm}, for the two emission wavelengths \SI{920}{\nm} and \SI{1090}{\nm}, with model fit \textbf{(b)} Signal enhancement in a \SI{12.1}{\cm} long waveguide at a wavelength of \SI{1090}{\nm} \textbf{(c)} Setup for signal enhancement measurements. FPC: fiber polarization controller, PBS: polarizing beam splitter, BS: beam splitter, OSA: optical spectrum analyzer. \textbf{(d)} Gain saturation measurement at a pump power of \SI{13.7}{\mW}.}
\end{figure*}

To demonstrate amplification of light in the neodymium-implanted waveguides, we measure signal enhancement and propagation losses.

Signal enhancement is measured by comparing the transmitted signal with and without pump on an optical spectrum analyzer (Fig.~\ref{fig:3-amplification}(c).
This is done for signal wavelengths of \SI{920}{\nm} and \SI{1090}{\nm}, close to the peak emission wavelengths.
Signal input powers are small (less than \SI{10}{\uW}) to avoid gain saturation.
Pump light at a wavelength of \SI{815}{\nm} is used.
Both signal and pump light propagate in the fundamental TE mode of the waveguide.

At a signal wavelength of \SI{920}{\nm}, up to \SI{4.5}{\dB} of signal enhancement is observed in a \SI{3.2}{\cm} long waveguide, and up to \SI{6.5}{\dB} in a \SI{5.9}{\cm} waveguide (Fig.~\ref{fig:3-amplification}(a)).
The gain at this wavelength is limited by the propagation losses.
There are both saturable propagation losses of \SI{0.2}{\dB\per\cm} due to absorption by neodymium ions and unsaturable scattering losses of \SI{0.7}{\dB\per\cm}.
Both are subtracted from the signal enhancement to obtain the gain, which is \SI{1.5}{\dB} in the \SI{3.2}{\cm} long amplifier and \SI{0.5}{\dB} in the \SI{5.9}{\cm} long amplifier.

Due to smaller losses and larger signal enhancement, larger gain is measured at a signal wavelength of \SI{1090}{\nm}.
Peak signal enhancement in the \SI{3.2}{\cm} long waveguide is \SI{6.5}{\dB}, and \SI{9.6}{\dB} in the \SI{5.9}{\cm} long waveguide (Fig.~\ref{fig:3-amplification}(a)).
After subtracting propagation losses of \SI{0.6}{\dB\per\cm}, the gain of these amplifiers is \SI{4.6}{\dB} and \SI{6.0}{\dB} respectively.

The largest signal enhancement was measured in a \SI{12.1}{\cm} long waveguide at a wavelength of \SI{1090}{\nm}.
There, we measure \SI{19.8}{\dB} of signal enhancement (Fig.~\ref{fig:3-amplification}(b)).

We determine the saturation power of the \SI{3.2}{\cm} long amplifier by measuring amplification of large signals at constant pump power (Fig.~\ref{fig:3-amplification}(d)).
From fitting $g = g_{ss} / (1 + P/ P_\text{sat})$ the data, the input signal power at which the gain drops to half its small-signal value is $P_\text{sat} = \SI{1.85}{\mW}$.

The propagation losses measured in ion-implanted waveguides are larger than for identical unimplanted waveguides.
A second sample from the same wafer prepared identically except for ion implantation was used to determine the propagation losses in waveguides without implantation damage by the cutback method.
At both signal wavelengths, this resulted in propagation losses of \SI{0.2}{\dB\per\cm}.
This is smaller than the values of \SI{0.6}{\dB\per\cm} to \SI{0.7}{\dB\per\cm} measured for the implanted waveguides.
This suggests that the waveguide is damaged by ion implantation, and that this damage was not fully repaired by high-temperature annealing.
This might be caused by performing the high-temperature annealing step after the waveguides have already been etched, which can cause deformation of thin waveguides (Supplementary note 2 of \cite{Kippenberg2026ThinNitrideLasers}).

The emission cross sections implied by the measured gain are smaller than those calculated assuming purely radiative decay.
The emission cross sections are determined by fitting a propagation and rate equation model for the gain (described in detail in the supplementary material) with only the emission cross section as a free parameter to the measured data.
There is good agreement for the shorter \SI{3.2}{\cm} waveguide, and worse agreement for the longer \SI{5.9}{\cm} waveguide.
This is likely due to the more complicated geometry with a less confined mode in the wider waveguide with bends, which is not taken into account by the model.
Because of this, we use the emission cross section determined from the fit to the signal enhancement by the \SI{3.2}{\cm} waveguide.
The resulting values for the emission cross section are \SI{1.0(3)e-20}{\cm^2} at a wavelength of \SI{920}{\nm} and \SI{1.7(4)e-20}{\cm^2} at a wavelength of \SI{1090}{\nm} (Fig.~\ref{fig:2-spectroscopy}(e)).
These values are significantly smaller than those calculated assuming purely radiative decay, which are \SI{3.4e-20}{\cm^2} at a wavelength of \SI{920}{\nm} and \SI{3.8e-20}{\cm^2} at a wavelength of \SI{1090}{\nm}.

The radiative lifetime that best matches the emission cross section values determined from signal enhancement measurements is approximately \SI{250}{\us}.
This radiative lifetime is obtained by using the Füchtbauer-Ladenburg relation (Eq.~\ref{eq:fuchtbauer-ladenburg-relation}) in reverse, choosing $\tau$ to match the measured values of $\sigma_e$.
This is similar to the value in other amorphous hosts, such as aluminum oxide (\SI{325}{\us}).
The large nonradiative decay rate may be inherent in silicon nitride as a host material, but it may also be caused by clustering, since the peak concentration of \SI{3e26}{m^{-3}} is close to concentrations at which concentration quenching is observed in different neodymium doped glasses \cite{Weber1984NdGlassConcentrationQuenching}.
In the latter case, the amount of nonradiative decay may be reduced by using multiple implantation energies to achieve the same effective number of ions at a lower peak concentration.

\subsection{Lasing in a microring resonator}

\begin{figure*}
\includegraphics{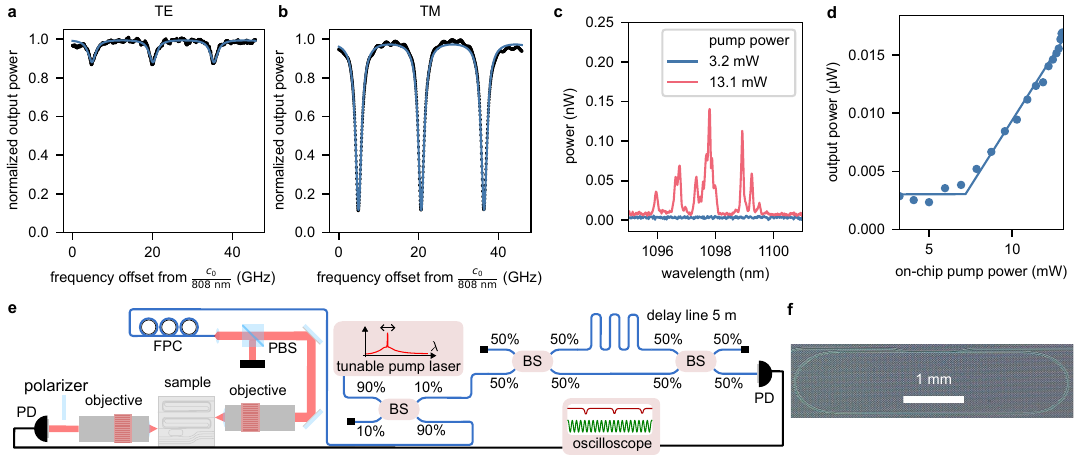}
\caption{\label{fig:4-lasing}\textbf{(a)}, \textbf{(b)} Pump transmission through the bus waveguide for TE and TM polarized light \textbf{(c)} Spectrum out of the bus waveguide at two different pump powers \textbf{(d)} Output power in the spectral region \SI{1095}{\nm} to \SI{1101}{\nm} for different pump powers \textbf{(e)} Setup for measuring pump absorption and lasing. FPC: Fiber polarization controller, BS: fiber beam splitter, PBS: polarizing beam splitter, PD: photodetector \textbf{(f)} Microscope picture of the Euler bend microring resonator and the bus waveguide.}
\end{figure*}

We demonstrate a microring laser at a wavelength of around \SI{1100}{\nm}.
It consists of a racetrack microring resonator (Fig.~\ref{fig:4-lasing}(f)), which is coupled to a bus waveguide.
The coupling at the pump wavelength is critical for the TM0 mode but not for the TE0 mode.
This is determined by measuring the transmission spectrum of the bus waveguide with a tunable pump laser. A fiber Mach-Zehnder interferometer is used as a frequency reference (Fig.~\ref{fig:4-lasing}(e)).
The free spectral range of the resonances matches the value expected for the \SI{10}{\mm} perimeter resonator with a mode contained mostly in the silicon nitride, which is $\SI{15}{\GHz}$.
The transmission almost drops to zero on resonance for the TM0 mode of the bus waveguide, indicating critical coupling.
Therefore, the resonator is pumped using the TM0 mode of the bus waveguide, even though the TE0 mode would have better overlap with the ion distribution.

We observe threshold behavior both in the output power and the output spectrum.
As the pump power is increased, a change of the spectrum of the light out of the bus waveguide is observed: Below a threshold pump power of \SI{7}{\mW}, the smooth spontaneous emission spectrum (as in Fig.~\ref{fig:2-spectroscopy}(b)) is observed. At pump powers above the threshold, sharp peaks appear in the spectrum (Fig.~\ref{fig:4-lasing}(c)).
The threshold behavior is also observed in the total output power in the spectral region around these peaks (Fig.~\ref{fig:4-lasing}(d)).

The multiple peaks in the output spectrum correspond to multiple longitudinal modes of the microring resonator lasing.
Due to the limited resolution of the optical spectrum analyzer of \SI{0.1}{\nm} corresponding to \SI{25}{\GHz} at a wavelength of \SI{1100}{\nm}, the individual lasing resonances cannot be resolved in the spectrum of the output.
The gaps in the lasing spectrum could be caused by spatial holeburning or interaction with counterpropagating lasing modes.

The sharp lines in the output spectrum are TE-polarized.
We measure the output spectrum transmitted through a polarizer.
When orienting the polarizer to transmit only TM-polarized light, only the smooth amplified spontaneous emission (ASE) background is observed.
When rotating the polarizer to transmit only TE-polarized light, the sharp lasing lines appear.

From the threshold behavior of the spectrum, the power, and the polarization of the output, we conclude that the microring resonator is lasing.

Lasing in the microring resonator is only observed at a wavelength of \SI{1100}{\nm}, but not at \SI{920}{\nm}.
This is likely due to the smaller gain, but might also be influenced by the coupling of the available modes to the bus waveguide.

\section{Discussion}
\label{sec:discussion}

\begin{figure}
    \includegraphics{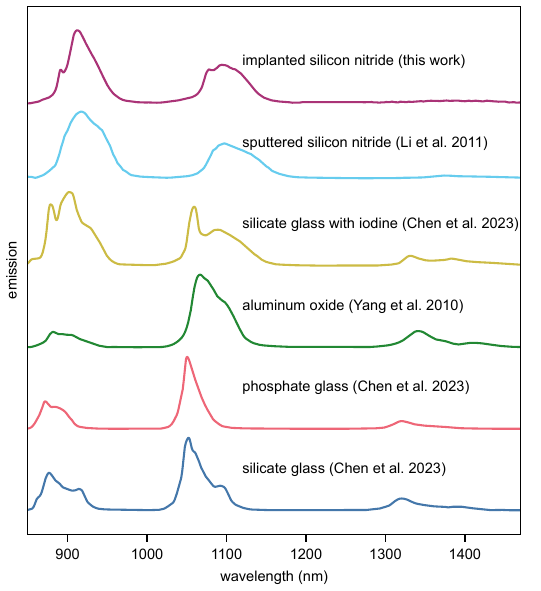}
    \caption{\label{fig:5-host-comparison}Comparison of the emission spectra of different amorphous hosts for $\mathrm{Nd}^{3+}$. Data for other hosts is reproduced from: Li et al. 2011 \cite{DalNegro2011SputteredNdSiN}, Chen et al. 2023 \cite{Hu2023NdFiberDopantEngineering}, Yang et al. 2010 \cite{Pollnau2010AlOxNd}.}
\end{figure}

The observed spectra are similar in structure to those in other amorphous hosts, silicate and phosphate glasses, as well as aluminum oxide (Fig.~\ref{fig:5-host-comparison}).
In contrast to these hosts, the emission at \SI{920}{\nm} is stronger than the emission around \SI{1100}{\nm}.
This is unusual, and in silica glass specific engineering of the co-dopants is necessary to achieve significant emission at \SI{900}{\nm} \cite{Hu2023NdFiberDopantEngineering}.
Additionally, there is almost no emission at \SI{1060}{\nm}, it is shifted to \SI{1100}{\nm}.

The observed FWHM emission bandwidths of \SI{10}{\THz} at a wavelength of \SI{920}{\nm} and \SI{15}{\THz} at a wavelength of \SI{1100}{\nm} are sufficiently large to amplify short pulses.
Especially the flat emission profile around \SI{1100}{\nm} is promising for this purpose.

The measured lifetime of the upper lasing level ${}^4 F_{3/2}$, \SI{85}{\us}, is significantly shorter than in other amorphous hosts:
In aluminum oxide, the lifetime is \SI{325}{\us} \cite{Pollnau2010AlOxNd}, in phosphate glass it is \SIrange{300}{400}{\us} \cite{Ropp1981NdPhosphateGlass}.
The value is comparable to the two time constants observed for the photoluminescence decay in sputtered silicon nitride, which cannot be described by a single exponential, of \SI{50}{\us} and \SI{140}{\us} \cite{DalNegro2011SputteredNdSiN}, although there are also reports of sputtered silicon nitride films with much shorter photoluminescence lifetime \cite{Gourbilleau2013SputteredSiN}.
The radiative lifetime of \SI{250}{\us} derived from the measured gain is much closer to the radiative lifetime in other amorphous hosts.

Notwithstanding the observed non-radiative decay, amplifiers with up to \SI{1.5}{\dB} of gain at a wavelength of \SI{920}{\nm} and \SI{6.0}{\dB} at a wavelength of \SI{1090}{\nm} were demonstrated.
The saturation power of the short amplifier at a wavelength of \SI{1090}{\nm} was shown to be \SI{1.85}{\mW}.
The amplifier gain and output power is limited by the per-length gain, losses, the amplifier length, and the available on-chip pump power, which only reached \SI{15}{\mW} for these amplifiers.
There is significant room for improving the per-length gain, as only a single implantation step was performed, resulting in an overlap of \SI{10}{\percent} of the light intensity with the ions.
Multiple implantation steps could improve this by a factor of 3 \cite{Kippenberg2026ThinNitrideLasers}.
Reports on erbium-implanted silicon nitride also suggest that the losses of the implanted waveguides can be reduced significantly \cite{Kippenberg2026ThinNitrideLasers}.
The demonstrated signal enhancement of \SI{19.8}{\dB} in a \SI{12.1}{\cm} waveguide indicates that more gain can be achieved in longer waveguides.
Finally, optimization of the edge couplers will allow for larger on-chip pump powers, further increasing the total gain. Laser diodes that provide multiple \SI{100}{\mW} in a single spatial mode at the required pump wavelength are available.
All of this suggests that neodymium-implanted silicon nitride waveguides could reach significantly larger gain and output power than demonstrated here.

Until now, neodymium-based lasing in micro resonators had been demonstrated in neodymium-doped silica microspheres \cite{Haroche1996MicrosphereLaser}, whereas in silicon nitride only stimulated emission in a microdisk resonator had been shown \cite{DalNegro2011SputteredNdSiN}.
The microring laser demonstrated in this work shows that the direct implantation of Nd into LPCVD silicon nitride provides the necessary gain overcoming the propagation losses needed for lasing. This was not achieved in sputtered Nd:SiNx microdisks. 

\section{Conclusion}

Neodymium-implanted silicon nitride has two broad emission bands at wavelengths of \SI{920}{\nm} (\SI{10}{\THz} bandwidth) and \SI{1100}{\nm} (\SI{15}{\THz} bandwidth).
It provides gain at these wavelengths upon pumping with light at a wavelength of \SI{815}{\nm}, even though we inferred that there is significant non-radiative decay from the confocal and waveguide gain measurements.
The demonstration of gain and lasing in neodymium-implanted silicon nitride shows that it is possible to integrate rare-earth-based gain at wavelengths near \SI{1.0}{\um} on established silicon nitride nanophotonic platforms, as has already been demonstrated using erbium-implanted silicon nitride waveguides around \SI{1.5}{\um}.

This approach enables the integration of optical gain at wavelengths of \SI{920}{\nm} and \SI{1100}{\nm} into existing ultralow-loss silicon nitride platforms, opening opportunities for applications in metrology, sensing, and quantum technologies. With further optimization, these waveguide amplifiers could enable tunable integrated lasers operating at both wavelengths, leveraging mature silicon nitride building blocks for filtering, multiplexing, and splitting.

\section{Methods}
\label{sec:methods}

\subsection{Fabrication}

Photonic integrated circuits were manufactured on LIGENTEC's \SI{200}{\mm} AN800 platform, based on low-pressure chemical vapor deposition (LPCVD) silicon nitride ($\mathrm{Si}_3\mathrm{N}_4$).
The waveguide stack consists of an \SI{800}{\nm}-thick
$\mathrm{Si}_3\mathrm{N}_4$ layer topped by an additional
\SI{200}{\nm}-thick $\mathrm{Si}_3\mathrm{N}_4$ layer. The two silicon nitride layers are separated by \SI{100}{\nm} of silicon oxide.
As illustrated in Fig.~\ref{fig:1-overview-and-fabrication},
the implantation was designed to confine the Nd$^{3+}$ ions
to the upper \SI{200}{\nm} thick layer, reducing the implantation energy
required and thus limiting implantation-induced damage
\cite{Kippenberg2026ThinNitrideLasers}.
The calculated overlap of the fundamental TE-polarized mode
with the ion distribution is \SI{10}{\percent} at a wavelength of \SI{1090}{\nm}, \SI{12}{\percent} at \SI{920}{\nm}, and \SI{13}{\percent} at \SI{815}{\nm}. The standard silicon oxide upper cladding of the AN800 platform was intentionally omitted to provide direct access to the silicon nitride
surface during implantation.

Following fabrication, the reticles were separated by stealth dicing and subsequently implanted with neodymium ions. A single neodymium implantation step was used to minimize implantation-induced damage to the waveguides. Additionally, oxygen co-doping was introduced to promote the optical activation of Nd$^{3+}$ ions, motivated by previous reports of enhanced rare-earth luminescence in oxygen-containing host materials \cite{Kenyon2002RareEarthOptoelectronics}. The neodymium implantation was performed at an energy of \SI{200}{\keV} and a fluence of \SI{1e15}{\cm^{-2}}.

Following implantation, the sample underwent ICP-PCVD silicon oxide deposition of 500 nm to prevent waveguide damage during thermal annealing at \SI{1000}{\celsius} for \SI{1}{\hour} in a nitrogen atmosphere.

\subsection{Lifetime measurement}

The lifetime of the ${}^4 F_{3/2}$ level is measured by observing the decay of the photoluminescence coupled to the waveguide on a silicon photodetector after turning off the \SI{815}{\nm} pump laser (Fig.~\ref{fig:2-spectroscopy}(d)).
The pump light is filtered out in the collection path using an \SI{850}{\nm} wavelength longpass filter.
The output of the laser is modulated with an acousto-optical modulator, which has a \SI{50}{\ns} fall time.
The bandwidth of the photodetector and the transimpedance amplifier is \SI{260}{\kHz}, corresponding to a fall time of \SI{4}{\us}, significantly shorter than the measured lifetime.
Multiple decay traces are averaged using an oscilloscope.
An exponential function with background is fitted to the data, $f(t) = A e^{-t/\tau} + B$.

The responsivity of the silicon photodetector is \SI{60}{\percent} at a wavelength of \SI{920}{\nm}, but drops to \SI{15}{\percent} at a wavelength of \SI{1100}{\nm}.
This means that the observed signal is dominated by the emission on the ${}^{4}\mathrm{F}_{3/2} \to {}^{4}\mathrm{I}_{9/2}$ transition, and that the emission on the ${}^{4}\mathrm{F}_{3/2} \to {}^{4}\mathrm{I}_{11/2}$ transition is mostly neglected.
Since the amount of spontaneous emission into the waveguide is proportional to the population in the ${}^{4}\mathrm{F}_{3/2}$ state for both transitions, this does not affect the measured lifetime.

\subsection{Absorption spectrum measurement}

The propagation loss spectrum was measured using a pulsed broadband source (SuperK EXTREME, NKT Photonics) and an optical spectrum analyzer by the cutback method.
The transmission through waveguides with lengths between \SI{3}{\cm} and \SI{8}{\cm} was measured and a linear fit was used to extract the propagation loss at each wavelength.
The measurement was done for the fundamental TM mode.
The minimum propagation loss of \SI{0.7}{\dB\per\cm} is taken to be the background loss caused, for example, by scattering due to sidewall roughness, and the excess is attributed to absorption by the $\mathrm{Nd}^{3+}$ ions.

\subsection{Emission spectrum measurement}

To avoid a distortion of the emission spectrum by reabsorption and amplification in the waveguide, as well as wavelength-dependent coupling losses, the $\mathrm{Nd}^{3+}$-implanted silicon nitride film is excited with an \SI{815}{\nm} pump laser and the spectrum collected using a confocal microscope (Fig.~\ref{fig:2-spectroscopy}(b)).
A shortpass filter is used at the laser output to filter out emission tails of the pump laser at the neodymium emission wavelengths.
The light emitted by the film is coupled into a \SI{50}{\um} core multimode fiber and the spectrum is determined using a grating spectrometer with an InGaAs CCD camera (Andor Kymera 328i spectrograph).
The emission spectrum of a commercial Nd:YAG crystal is also recorded.
The known positions of the emission peaks of Nd:YAG are used to calibrate the wavelength axis of the spectrometer.
The known relative intensities are used to calibrate the intensity response of the spectrometer.

\subsection{Optical gain measurement}

Signal enhancement is measured with counterpropagating signal and pump light.
To couple both signal and pump light, microscope objectives are used, which can collect light emitted in any waveguide mode up to their numerical aperture (\num{0.65}).
All coupling losses occur due to mode mismatch at the input objective, while no coupling losses occur at the collecting output objective.
Because pump and signal are incident from opposite sides, the objective positions can be adjusted to allow for optimal incoupling efficiency for both pump and signal light.

Coupling losses at the input are then determined by cutback measurements.
At the pump wavelength sufficiently low optical powers are used so that the absorption is not saturated.

To ensure controlled polarization and power, polarizers are used behind the single mode fibers delivering the pump and signal light.
The power is monitored behind these polarizers.
The fibers are used to ensure that changes in laser power do not affect the pointing of the beams.

An optical spectrum analyzer is used to determine the transmitted signal power to distinguish broadband spontaneous emission power from signal power.
The power measured on the OSA is calibrated by measuring the power out of the objective with a power meter.
The transmission of the objectives is also measured at all relevant wavelengths and taken into account.

The measurement of the \SI{12.1}{\cm} long waveguide was optimized for maximal signal enhancement and no polarization control was used. Moreover, the coupling losses could not be determined due to the omission of the single mode fiber for this measurement. This differs from the measurement setup shown in \ref{fig:3-amplification}(c) in the omission of the single mode fiber and polarizer for the pump beam.

\FloatBarrier

\section*{Data availability}
The data supporting the findings of this study
are available from the corresponding author
upon reasonable request.

\begin{acknowledgments}

\subsection{Funding}
Arturo Fernández is funded by HORIZON EUROPE Marie Sklodowska-Curie Actions (MicrocombSys,101119968). This work is supported by the European Research Council (ERC) under the European Union’s Horizon Europe research and innovation program through the Starting Grant LASIQ, grant agreement No. 101078281.

\subsection{Author contributions}
A.F.G. and K.V.G. conceived the idea.  A.F.G. designed the $\mathrm{Si}_3\mathrm{N}_4$ wafer reticle and post-processed the wafers. A.F.G. and U.K. performed the neodymium implantation simulations. U.K. performed the ion implantation and R.H. performed data analysis. K.V.G., J.H., N.Z., and K.K. designed and performed the experiments, data analysis, and simulations. A.F.G., J.H., and K.V.G. wrote the manuscript with input from all co-authors. A.C., B.K. and K.V.G. supervised the project.

\end{acknowledgments}

We thank Maxime Delaey and Dirk Poelman for annealing the samples, and Nicolas Le Thomas for providing the pump laser.

\renewcommand{\bibsection}{\section*{\refname}}
\bibliographystyle{unsrtnat}
\bibliography{literature}

\clearpage
\onecolumngrid

\input{supplementary}

\end{document}

%% file: supplementary.tex
\section{Appendix}

\input{model}

%% file: model.tex
\subsection{Amplifier model}
\label{sec:apdx-amplifier-model}

Here, we describe the model we use for the waveguide amplifiers.
We only consider a single transverse mode of the electromagnetic field supported by the waveguide for each frequency of light (in the experiments we use the TE0 mode), and describe the light in the frequency domain by the spectral power propagating in this spatial mode.
The mode of the electromagnetic field in the waveguide is computed from the fabricated waveguide geometry with a finite difference eigenmode solver.
The intensity distribution is then given by the $z$-component of the Poynting vector, $I(x, y) = \frac{1}{2} \Re (\vec{E}(x, y) \times \vec{H}(x, y)^*)_z$.
This intensity distribution is used for calculating the interaction between the ions and the light.
For a spectral power $\tilde{P}(\nu)$ at a position along the waveguide, the spectral intensity is given by $\tilde{P}(\nu) = \int \tilde{I}(\nu, x, y) \dd{x} \dd{y}$.

The ions are modeled as follows: Absorption of a photon with a wavelength around \SI{800}{\nm} by an ion in the ground ${}^4 I_{9/2}$ state brings the ion into the short-lived ${}^4 F_{5/2}$ state, from where it very quickly decays into the long-lived ${}^4 F_{3/2}$ state. Absorption of light with a wavelength of around \SI{900}{\nm} brings the ion directly into the long-lived ${}^4 F_{3/2}$ state. From there it may decay to any of the four levels ${}^4 I_{15/2}$, ${}^4 I_{13/2}$, ${}^4 I_{11/2}$, and ${}^4 I_{9/2}$ by spontaneous or stimulated emission, or through non-radiative decay.
The populations of the short-lived levels are not included in the model.
Only the populations of the ground state ${}^4 I_{9/2}$ and of the long-lived upper lasing level ${}^4 F_{3/2}$ are kept track of.
For the ground state ${}^4 I_{9/2}$, only absorption is modeled, and for the lasing level ${}^4 F_{3/2}$ only emission is modeled.

With these assumptions, propagation of light along the waveguide is described by an equation for the rate of change of the spectral power $\dv{\tilde{P}(\nu)}{z}$ in the selected mode along the waveguide. The equation includes the following terms (here they are given for the forward-propagating modes, the sign is inverted for the backward-propagating modes):
\begin{itemize}
    \item Propagation losses unrelated to Nd-absorption, which are always the same independent of power, $-\alpha_0(\nu) \tilde{P}(\nu)$.
    \item Absorption by the neodymium ions in the ground state, $-\Gamma(\nu) \sigma_a(\nu) N_1 \tilde{P}(\nu)$. The product of the peak concentration of ground-state ions in the waveguide, $N_1$, and the absorption cross section of the ions $\sigma_a(\nu)$, gives the attenuation coefficient of light through a medium in which the concentration of ground-state ions is $N_1$ everywhere. Because the ions are concentrated in one part of the waveguide, the actual attenuation coefficient is smaller in magnitude, and this is taken into account by the overlap factor $\Gamma(\nu)$ of the ion distribution with the intensity.
    This overlap factor is defined as
    \begin{equation}
        \Gamma(\nu) = \frac{\int I(x, y) N_\text{Nd}(x, y) \dd{x} \dd{y}}{P N_\text{max}}
    \end{equation}
    \item Stimulated emission by the neodymium ions in the excited state, $\Gamma(\nu) \sigma_e(\nu) N_1 \tilde{P}(\nu)$. This term completely mirrors the absorption term, but the product of excited state concentration and emission cross section gives an amplification coefficient. The overlap factor is defined in the same way.
    \item Spontaneous emission by the Nd ions: The rate at which spectral power is added along the waveguide due to spontaneous emission of light into the waveguide mode by decay of the excited state is $\Gamma(\nu) \sigma_e(\nu) N_2 h \nu$ \cite{Simpson1999EDFABook}.
\end{itemize}

The concentrations of ions in the ground state $N_1$ and in the excited state $N_2$ are determined by the intensity at the location of the ions by rate equations which include the following terms:
\begin{itemize}
    \item Absorption rate: $\int \sigma_a(\nu) \tilde{\phi}(\nu) \dd{\nu}$
    \item Stimulated emission rate: $\int \sigma_e(\nu) \tilde{\phi}(\nu) \dd{\nu}$
    \item Spontaneous emission and non-radiative decay rate: $\frac{1}{\tau}$
\end{itemize}
These rates depend on the spectral photon flux $\tilde{\phi}(\nu) = \tilde{P}(\nu) / (A(\nu) h \nu)$.
In principle, the intensity varies across the ion distribution, but we use the average intensity at the ions, i.e. we define an effective area $A$ such that $P / A$ is equal to the average of the mode intensity weighed by the ion distribution.

Putting everything together, the complete set of propagation equations considering light at any frequency propagating forward and backward are
\begin{align}
    \pm \dv{\tilde{P}_\pm(\nu)}{z} &= -\alpha_0(\nu) \tilde{P}_\pm(\nu)\\
    &- \Gamma(\nu) \sigma_a(\nu) N_1 \tilde{P}_\pm(\nu)\\
    &+ \Gamma(\nu) \sigma_e(\nu) N_2 \tilde{P}_\pm(\nu)\\
    &+ \Gamma(\nu) \sigma_e(\nu) N_2 h \nu
\end{align}
and
\begin{align}
    \dv{N_1}{t} &= \frac{1}{\tau} N_2 + \int (\sigma_e N_2 - \sigma_a N_1) \tilde{\phi}(\nu) \dd{\nu}\\
    \dv{N_2}{t} &= -\dv{N_1}{t}\\
    N_1 + N_2 &= N_\text{max}
\end{align}
with the solution in steady state for $\dv*{N_1}{t} = 0$
\begin{align}
    N_1 &= N_\text{max} \frac{1/\tau + \int \sigma_e(\nu) \tilde{\phi}(\nu) \dd{\nu}}{1/\tau + \int (\sigma_e(\nu) + \sigma_a(\nu)) \tilde{\phi}(\nu) \dd{\nu}}
\end{align}

In the simple case of monochromatic counterpropagating pump and signal light, i.e. $\tilde{\phi}_+(\nu) = \phi(\nu_s) \delta(\nu - \nu_s)$ and $\tilde{\phi}_-(\nu) = \phi(\nu_p) \delta(\nu - \nu_p)$ for $\phi(\nu) = P(\nu) / (h \nu A(\nu))$, and neglecting spontaneous emission, we obtain the following equations:
\begin{align}
    \dv{P_+(\nu_p)}{z} &= (-\alpha_0(\nu_p) - \Gamma(\nu_p) \sigma_a(\nu_p) N_1 + \Gamma(\nu_p) \sigma_e(\nu_p) N_2) P_+(\nu_p)\\
    \dv{P_-(\nu_s)}{z} &= -(-\alpha_0(\nu_s) - \Gamma(\nu_s) \sigma_a(\nu_s) N_1 + \Gamma(\nu_s) \sigma_e(\nu_s) N_2) P_-(\nu_s)\\
    N_1 &= N_\text{max} \frac{1/\tau + \sigma_e(\nu_p) \phi(\nu_p) + \sigma_e(\nu_s) \phi(\nu_s)}{1/\tau + \sigma_e(\nu_p) \phi(\nu_p) + \sigma_e(\nu_s) \phi(\nu_s) + \sigma_a(\nu_p) \phi(\nu_p) + \sigma_a(\nu_s) \phi(\nu_s)}
\end{align}
For initial conditions $P_+(\nu_p)|_{z=0} = P_{p,0}$ and $P_+(\nu_s)|_{z=L} = P_{s,0}$ for an amplifier of length $L$ with pump and signal powers $P_{p,0}$ and $P_{s,0}$ on chip, the resulting boundary value problem may be solved numerically.

Two aspects that are not taken into account in the model are any inhomogeneity in the absorption and emission spectra of the ions that is not hidden by cross-relaxation between the ions, and excited-state absorption.